\documentclass[a4paper,11pt]{article}
\usepackage{pos}
\usepackage{braket}
\usepackage{amsmath}
\usepackage{amssymb}
\usepackage{multicol}
\usepackage{slashed}
\usepackage{mathtools}
\usepackage{float}
\usepackage{dsfont}
\usepackage{caption}
\usepackage{subcaption}
\usepackage{lipsum}

\usepackage[utf8]{inputenc}\DeclareUnicodeCharacter{2212}{-}
\usepackage[normalem]{ulem}

\title{Towards an Effective String Theory for the flux tube}

\newcommand{\be}{\begin{equation}}
\newcommand{\ee}{\end{equation}}
\newcommand{\bea}{\begin{eqnarray}}
\newcommand{\eea}{\end{eqnarray}}

\author*[a]{Andreas Athenodorou}
\author[b,c]{Sergei Dubovsky}
\author[b,d]{Conghuan Luo}
\author[e,f]{Michael~Teper}

\affiliation[a]{Computation-based Science and Technology Research Center, The Cyprus Institute, Cyprus}
\affiliation[b]{Center for Cosmology and Particle Physics, Department of Physics, New York University, New York, NY, 10003, USA}
\affiliation[c]{Institute for Advanced Study, Princeton, NJ 08540, USA}
\affiliation[d]{Department of Physics and Astronomy, University of Southern California, Los Angeles, CA, 90089, USA}
\affiliation[e]{Rudolf Peierls Centre for Theoretical Physics, University of Oxford, Parks Road, Oxford OX1 3PU, UK}
\affiliation[f]{All Souls College, University of Oxford, High Street, Oxford OX1 4AL, UK}

\abstract{The quest to develop an effective string theory capable of describing the confining flux tube has been a longstanding objective within the theoretical physics community. Recent lattice results indicate that the low-lying spectrum of the flux tube in both three and four dimensions can be partially described by the Nambu-Goto string with minor deviations. However, several excitation states exhibit significant corrections that have remained unexplained until recently. Recent advancements suggest that a Thermodynamic Bethe Ansatz (TBA) analysis, expanded in both $1/R \sqrt{\sigma}$ and the softness of phonons i.e. $p/\sqrt{\sigma}$, can lead to a robust effective string theory for the flux-tube with length $R$. Furthermore, lattice data points to the existence of an axion field on the world-sheet of the flux-tube, implying that an Axionic String Ansatz (ASA) should accompany the Nambu-Goto framework. We will provide compelling evidence in these proceedings that this approach can closely approximate the flux tube data. We will demonstrate this by comparing results obtained for the spectrum of the closed $SU(N_c)$ flux-tube extracted using lattice techniques in four dimensions.}

\FullConference{%
  The 41st International Symposium on Lattice Field Theory (LATTICE2024),\\
  28 July - 3 August 2024 \\
  Liverpool, UK 
}

\begin{document}
\maketitle

\section{Introduction}
\label{sec:introduction}
In confining gauge theories like $SU(N_c)$ gluodynamics, the energy of a gauge field sourced by color charges forms narrow flux tubes, also known as confining strings. Understanding the worldsheet dynamics of these strings is crucial for comprehending the insights of color confinement, which remains unresolved despite numerous theoretical efforts.

To leverage lattice data, one needs a theoretical framework to relate the measured excitation spectrum of confining flux tubes to the fundamental parameters of the worldsheet theory. This can be approached by either starting with a candidate worldsheet theory and comparing its excitation spectrum to lattice data, or extracting worldsheet dynamics directly from lattice data in a model-independent manner.

The first approach works best when the confining string is long, where the dynamics are universal and described by the low-energy theory of transverse Goldstone modes (phonons), approximated by the Nambu-Goto action with subleading corrections. These proceedings focus on the excitations of torelons, in other words closed confining strings, that wind around a compact spatial direction. Traditionally, their excitation spectrum is calculated using a perturbative expansion in powers of \(\ell_s^2/R^2\), where \(\ell_s^{-2} = \sigma\) is the string tension.

However, high precision measurements of torelon excitations in $D=2+1$ and $D=3+1$ Yang-Mills theory reveal a discrepancy: the ground state energy matches universal perturbative predictions, but the agreement for excited states is less accurate. Surprisingly, many excited state energies align well with the GGRT spectrum~\cite{goddard1973quantum} (Goddard-Goldstone-Rebbi-Thorn), even at large \(\ell_s^2/R^2\). This suggests a need for a perturbative approach with better convergence, achieved through the Thermodynamic Bethe Ansatz (TBA), which connects the finite volume spectrum calculation into perturbatively computing worldsheet scattering amplitudes with finding the corresponding finite volume spectrum.

Lattice studies in 2+1 dimensional $SU(N_c)$ gauge theories show no additional massive string excitations, but in 3+1 dimensions "anomalously behaving" pseudoscalar states deviate significantly from the GGRT spectrum. TBA application reveals these states arise from a massive pseudoscalar resonance coupled to the Goldstone modes (worldsheet axion). The leading axion-Goldstone-Goldstone coupling from lattice data matches that in an integrable model, despite the worldsheet theory not being integrable.

Both 3D and 4D flux tubes exhibit a matter content compatible with possible integrable models, though exact integrability is absent. This observation supports the Axionic String Ansatz (ASA), proposing that integrability might be restored at high energies for some observables. An essential test for ASA is checking for no additional massive states on the worldsheet with higher torelon excitations.

The proceedings, adopting the structure of the actual talk~\cite{athenodorou_2024_13270842}, is organized as follows: Section~\ref{sec:Effective_String_Theory} reviews the effective string theory and ASA. Section~\ref{sec:lattice} discusses lattice methods and conventions. Sections~\ref{sec:results}-\ref{sec:sectionq1} use the TBA method to compute spectra of up to two-phonon excitations, employing the \(T\bar{T}\) technique for higher excitations. Finally, Section~\ref{sec:conclusions} concludes with the finding that all observed states can be described using Goldstones and the axion, with \(T\bar{T}\) deformation accurately modeling their interactions.

\section{Effective Theory of Long Strings}
\label{sec:Effective_String_Theory}

Long strings, described by translational Goldstone bosons with non-linearly realized Poincaré symmetry, have low-energy excitations that can be captured by the effective action $
S = -\int d^2 \sigma \sqrt{-\det h_{\alpha\beta}} \left[ \ell_s^{-2} + \gamma \ell_s^2 \mathcal{R}^2 + \dots \right],$ where \(\ell_s^{-2}\) is the string tension, \(\gamma\) is a Wilson coefficient, and \(\mathcal{R}\) is the scalar curvature. The first non-universal term arises at \(\mathcal{O}(\ell_s^2)\), making the Nambu–Goto action sufficient for universal predictions up to one-loop order.

For strings compactified on a circle of circumference \(R\), the perturbative \(\ell_s/R\) expansion can be used to calculate the torelon spectrum. While the leading non-universal contribution scales as \(\ell_s^6/R^7\) for long strings, lattice simulations show that many torelon states fit well with the GGRT (Goddard-Goldstone-Rebbi-Thorn) spectrum~\cite{goddard1973quantum,arvis1983exactqq}:

\begin{eqnarray}
E_{\text{GGRT}}(N_L, N_R) = \sqrt{\frac{4\pi^2(N_L - N_R)^2}{R^2} + \frac{R^2}{\ell_s^4} + \frac{4\pi}{\ell_s^2}\left(N_L + N_R - \frac{D-2}{12}\right)}.
\label{sec:GGRT_formula}
\end{eqnarray}

\subsection{Flux Tube Spectrum from Worldsheet Scattering}
\label{subsec:TBA}

An improved perturbative approach involves two steps: first, calculating phonon scattering amplitudes in the worldsheet effective theory using $p\ell_s$ as a small parameter; second, relating these amplitudes to the finite volume spectrum, potentially non-perturbatively. At tree level, phonon scattering is integrable with a flavor-independent $2 \to 2$ phase shift~\cite{Dubovsky:2012wk} $e^{2i \delta(s)} = e^{is \ell_s^2 /4}$, where $s$ is the Mandelstam invariant. This integrability allows employing TBA~\cite{zamolodchikov1990thermodynamic,Dorey:1996re} to exactly compute the finite volume spectrum.

Higher-order corrections, up to ${\cal O}(\ell_s^4)$, maintain the absence of particle production. The corresponding scattering phase shifts at $D=4$ are~\cite{Dubovsky:2012wk} for symmetric (tensor-$2^{\pm}$), antisymmetric (pseudoscalar-$0^{--}$) and singlet (scalar-$0^{++}$):
{\footnotesize\begin{equation}
    2 \delta_{sym} = \frac{\ell_s^2 s}{4} -\frac{11 \ell_s^4 s^2}{192 \pi}  +O(s^3), \quad 2 \delta_{anti} = \frac{\ell_s^2 s}{4} + \frac{11 \ell_s^4 s^2}{192 \pi}  + O(s^3), \quad 2 \delta_{sing} = \frac{\ell_s^2 s}{4} + \frac{11 \ell_s^4 s^2}{192 \pi} + O(s^3) \,,
    \label{phase_shift}
\end{equation}}
respectively, with the leading term matching the tree-level integrable phase shift. The subleading term, known as the Polchinski-Strominger (PS) amplitude, represents one-loop scattering and is universal up to this order.

Notably, the one-loop phase shifts for the scalar and pseudoscalar channels are identical, leading to reflectionless scattering in the helicity basis $a_{l(r) \pm}^{\dagger}=a_{l(r) 2}^{\dagger} \pm i a_{l(r) 3}^{\dagger} \,,$ where $a_{l(r)}^\dagger$ are creation operators for left-(l) and right-(r) moving phonons.

The TBA for massless, reflectionless excitations involves quantization conditions around the circle for each string particle:
{\footnotesize
\begin{equation}
\begin{aligned}
    p_{l i} R+\sum_j 2 \delta_{a_i a_j}\left(p_{l i}, p_{r j}\right)-i \sum_b \int_0^{\infty} \frac{d q}{2 \pi} \frac{d 2 \delta_{a_i b}\left(i p_{l i}, q\right)}{d q} \ln \left(1-e^{-R \epsilon_r^b(q)}\right)=2 \pi N_{li} \,, \\
    p_{r i} R+\sum_j 2 \delta_{a_j a_i}\left(p_{r i}, p_{l j}\right)+i \sum_b \int_0^{\infty} \frac{d q}{2 \pi} \frac{d 2 \delta_{b a_i}\left(-i p_{r i}, q\right)}{d q} \ln \left(1-e^{-R \epsilon_l^b(q)}\right)=2 \pi N_{ri} \,,
\end{aligned}
    \label{momentum_quantization}
\end{equation}}
where $a_i$, $a_j$ label left and right-moving excitations, and $b$ labels all particle species on the (1+1)-dimensional system. Neglecting winding corrections reduces TBA to the Asymptotic Bethe Ansatz (ABA), effectively resumming classical non-linearities. It is worth noting that for two-particle states the ABA is nothing but the well-known Luscher relation between the finite volume spectrum and the scattering amplitude \cite{Luscher:1990ux}.
Winding corrections in (\ref{momentum_quantization}) and (\ref{pseudoenergy}) depend on pseudo-energies $\epsilon_{l(r)}^a$, satisfying:
{\footnotesize
\begin{equation}
\begin{aligned} 
\epsilon_l^a(q) & =q+\frac{i}{R} \sum_i 2 \delta_{a b_i}\left(q,-i p_{r i}\right)+\frac{1}{2 \pi R} \sum_b \int_0^{\infty} d q^{\prime} \frac{d 2 \delta_{a b}\left(q, q^{\prime}\right)}{d q^{\prime}} \ln \left(1-e^{-R \epsilon_r^b\left(q^{\prime}\right)}\right) \,, \\ \epsilon_r^a(q) & =q-\frac{i}{R} \sum_i 2 \delta_{b_i a}\left(q, i p_{l i}\right)+\frac{1}{2 \pi R} \sum_b \int_0^{\infty} d q^{\prime} \frac{d 2 \delta_{b a}\left(q, q^{\prime}\right)}{d q^{\prime}} \ln \left(1-e^{-R \epsilon_l^b\left(q^{\prime}\right)}\right) \,.
\end{aligned}
\label{pseudoenergy}
\end{equation}}

The state's energy is given by:
{\footnotesize
\begin{equation}
    \Delta E=\sum_i p_{l i}+\sum_i p_{r i}+\frac{1}{2 \pi} \sum_a \int_0^{\infty} d q \ln \left(1-e^{-R \epsilon_l^a(q)}\right)+\frac{1}{2 \pi} \sum_a \int_0^{\infty} d q \ln \left(1-e^{-R \epsilon_r^a(q)}\right) \,,
    \label{energy_TBA}
\end{equation}}
where $\Delta E$ excludes the linear term $R/\ell_s^2$ from the worldsheet cosmological constant. Using only the leading phase shift in~\eqref{phase_shift}, the TBA simplifies to algebraic equations, yielding the GGRT formula~\eqref{sec:GGRT_formula}. Thus, in the TBA framework, the GGRT spectrum emerges as the leading approximation, explaining its efficacy in fitting many string states. The PS phase shift introduces universal corrections, breaking some degeneracies present at leading order.

Incorporating higher-order corrections into the full TBA is challenging both technically and conceptually. However, one can often approximate by neglecting higher-order phase shift corrections in winding terms, justified by their exponential suppression at high momenta. For example, in a two-phonon state, pseudoenergies become linear in momenta:
$\epsilon_{l(r)}^1(q) = \epsilon_{l(r)}^2(q) = c_{l(r)} q \,,$ leading to simplified TBA equations:
\begin{equation}
\begin{aligned}
    c_l = 1 + \frac{p_r \ell_s^2}{R} - \frac{\pi \ell_s^2}{6 c_r R^2}  \,, &\quad
    c_r = 1 + \frac{p_l \ell_s^2}{R} - \frac{\pi \ell_s^2}{6 c_l R^2}  \,, \\
    p_l R + 2 \delta(p_l, p_r) - \frac{\pi \ell_s^2 p_l}{6c_r R} = 2\pi N_L \,, &\quad
    p_r R + 2 \delta(p_r, p_l) - \frac{\pi \ell_s^2 p_r}{6c_l R} = 2\pi N_R \,,
    \label{TBA_equations}
\end{aligned}
\end{equation}
with the energy~\eqref{energy_TBA} expressed as $\Delta E = p_l + p_r - \frac{\pi}{6 R c_l} - \frac{\pi}{6 R c_r} \,.$

This approach aligns closely with numerical solutions of the full TBA system~\cite{Dubovsky:2014fma}, confirming the minimal UV sensitivity of winding corrections. Moreover, these equations allow extraction of scattering phase shifts as functions of $p_l,p_r$ from the spectrum of two-phonon excitations of a confining flux tube.

\subsection{The Axionic String Ansatz (ASA)}
\label{subsec:ASA}

Universal TBA predictions accurately describe a large sector of confining string excitations, as measured with high precision in both $D=4$ and $D=3$ \cite{Athenodorou:2010cs,Athenodorou:2011rx,Athenodorou:2018sab,Athenodorou:2022pmz}. As expected, most states show increasing deviations from these predictions at shorter string lengths $R$, suggesting the influence of non-universal corrections. The TBA analysis, in particular, allows for the determination of the first non-trivial Wilson coefficient for $D=3$ confining strings~\cite{Dubovsky:2014fma,Chen:2018keo}. However, some $D=4$ states exhibit significant deviations from the universal TBA predictions even at relatively large $R$, hinting at the presence of an additional massive excitation on the string worldsheet, referred to as the string axion \cite{Dubovsky:2013gi,Dubovsky:2014fma}.

The leading-order interactions between the worldsheet axion and the transverse Goldstone modes are described by the following action:
\begin{equation}
    S_\phi= \int d^2 \sigma \sqrt{-h} \left( -\frac{1}{2} (\partial \phi)^2 - \frac{1}{2} m^2 \phi^2 + \frac{Q_{\phi}}{4} h^{\alpha \beta} \epsilon_{\mu \nu \lambda \rho} \partial_\alpha t^{\mu \nu} \partial_\beta t^{\lambda \rho} \phi \right) \,,
    \label{axion_interaction}
\end{equation}
where $\phi$ is the pseudoscalar worldsheet axion, and $t^{\mu \nu}=\frac{\epsilon^{\alpha \beta}}{\sqrt{-h}} \partial_\alpha X^\mu \partial_\beta X^\nu$. The coupling constant $Q_\phi$ and axion mass $m$ can be extracted from Monte Carlo data of the 4d $SU(3)$ Yang-Mills confining flux tube spectrum using the TBA analysis $Q_{\phi} \approx 0.38 \pm 0.04, \quad m \approx 1.85_{-0.03}^{+0.02} \ell_s^{-1}$. Interestingly, this coupling aligns (within error bars) with the value \cite{Dubovsky:2015zey} $Q_{\text{integrable}} = \sqrt{7\over 16 \pi} \approx 0.373$ which corresponds to the coupling in an integrable theory of $D=4$ Goldstones and a massless axion.

\section{Lattice Calculation}
\label{sec:lattice}
We adopted standard Lattice techniques. Our calculations are done on a hypercubic periodic lattice of size $L_x L_{\perp}^2 L_t$ with lattice spacing $a$, focusing on flux tubes wound once around the $x$ direction with physical size $R = a L_x$.

We use the Wilson plaquette action $\beta S = \beta \sum_p \left\{1-\frac{1}{N} \text{ReTr} U_p\right\},$ where $U_p$ is the path-ordered product of link matrices around plaquette $p$ and $\beta = \frac{2N}{g^2}$ is the inverse coupling on the lattice. The 't Hooft large $N_c$ limit is approached by keeping $g^2N$ fixed, implying $\beta \propto N_c^2$ to maintain constant lattice scale $a$. The $\beta$ values used in the investigation are the following: for $N_c=3$, $\beta=6.338$ with $a\sqrt{\sigma}=0.12902(15)$ and $\beta=6.0625$ with $a\sqrt{\sigma}=0.19489(16)$, for $N_c=5$, $\beta=18.375$ with $a \sqrt{\sigma}=0.13047(25)$ and $\beta=17.630$ with $a\sqrt{\sigma}=0.19707(30)$ as well as for $N_c=6$, $\beta=25.550$ with $a \sqrt{\sigma}=0.20142(27)$.

The Torelon energies are extracted using the standard method of constructing a basis of operators that project irreducibly onto the quantum numbers characterizing the states of interest. This involves calculating the correlation matrix of these operators at various Euclidean time separations and then solving the Generalized Eigenvalue Problem (GEVP). All shapes of spatial Polyakov lines used to build these operators are detailed in Ref~\cite{Athenodorou:2024loq}. 

To construct the operators, we take linear combinations of rotations about the principal string axis and reflections of these shapes over both the transverse and longitudinal planes. The relevant quantum numbers that describe our operators and the string states, include the angular momentum \(|J|\) modulo 4, the transverse parity \(P_{\perp}\), the longitudinal parity \(P_{||}\), and the longitudinal momentum \(p = \frac{2 \pi q}{L_x}\). For simplicity we denote the states as $|J|^{P_{\perp} P_{||}}$.

\section{Results: Spectrum from the Thermodynamic Bethe Ansatz}
\label{sec:results}

In this section, we theoretically compute the $D=4$ Yang-Mills low-lying flux tube spectrum using TBA and $T\bar{T}$ deformation techniques and compare it to our Lattice results. We confirm that all accessible low-lying excited states of the confining flux tube in lattice simulations can be described by an effective long string theory with one massive pseudoscalar, with no additional low-energy matter. Specifically, we analytically compute the spectrum corresponding to string states with massive excitations and up to two phonon excitations, using $2 \to 2$ phonon scattering phase shifts from the proposed effective action, fitted with a small set of parameters. Moreover, in many cases we verify that the ``free'' $T\bar{T}$ dressing provide quantitatively good description of the axion-axion and axion-phonon interactions. 

We focus on the gauge group $SU(3)$ at coupling $\beta = 6.338$, as it provides the highest statistics for high-level excited states. Our data shows that the $N_c$ and lattice spacing dependence are insignificant, and do not affect our conclusions.

\section{Zero longitudinal momentum and the Worldsheet axion}
\label{spin0}

\begin{figure}[htb]
\scalebox{0.6}{\includegraphics[width=0.77\textwidth]{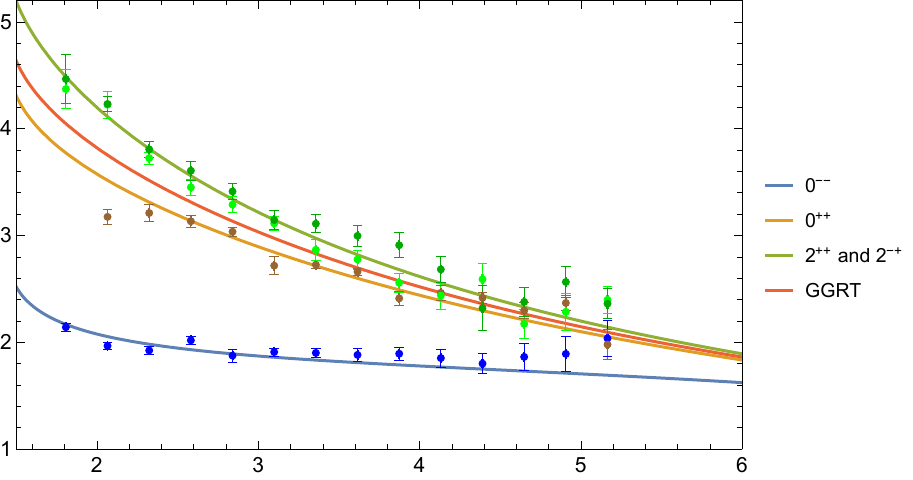}\put(-380,160){ \small $\Delta E\ell_s$}\put(-50,0){\small $R/\ell_s$}} \scalebox{0.6}{\includegraphics[width=0.65\textwidth]{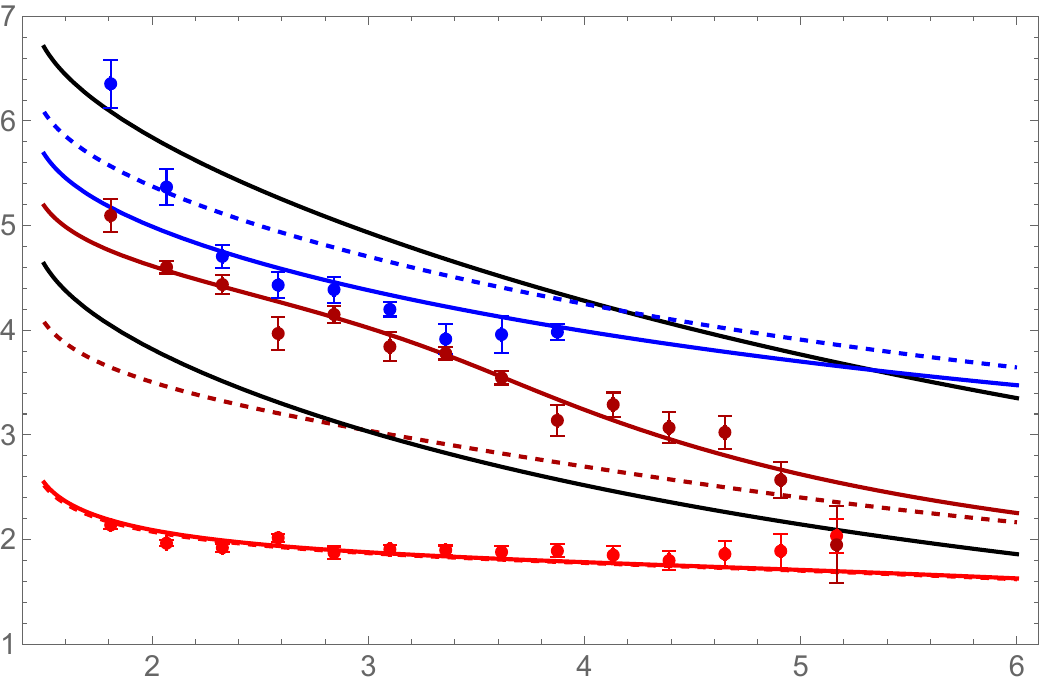}\put(-330,160){ \small $\Delta E\ell_s$}\put(0,0){\small $R/\ell_s$}}
\centering
\caption{{\underline{Left Panel:}} The energy at the level $N_L=N_R=1$. Solid lines represent the theoretical predictions of the spectrum of the two-phonon excitation states deriving from the $2 \to 2$ phonon anti-symmetric(blue), singlet(brown) and symmetric(green) scattering channels with the inclusion of a massive pseudo-scalar field, using ABA. Light green dots represent $2^{++}$ states, and dark green dots $2^{-+}$ states. \underline{Right Panel}: The energy of the ground (bright red), the first excited states (dark red) and the second excited states (blue) as a function of the string length $R$ for the $0^{--}$ sector; the black lines correspond to the GGRT predictions of the first ($N_L=N_R=1$) and second ($N_L=N_R=2$) excited states. The energies above are deducted by the linear piece $\Delta E = E - R/l^2_s$.}
\label{fig:p0n1}
\end{figure}

The absolute ground state of the confining flux tube with quantum numbers $0^{++}$, lacks phonon excitations and is well described by the GGRT spectrum and the $\ell_s/R$ expansion. This state allows for the extraction of the string tension using the GGRT formula giving $a^2 \ell^{-2} = 0.01665(4)$. Ground states in the $0^{--}$, $2^{++}$, $2^{-+}$ sectors, and the first excited state in $0^{++}$ show significant deviations from GGRT predictions, particularly in the $0^{--}$ state. These deviations can be explained by the effective string theory, including a massive pseudoscalar field. The pseudoscalar acts as a resonance on the string world-sheet, influencing phase shifts in phonon scattering and altering the energy spectrum of excited states. This resonance introduces the mass and the axion coupling as fitting parameters. By performing the fit we obtain $m = 1.812(16) \ell_s^{-1}, \; Q_{\phi} = 0.365(5).$ Theoretical predictions align with lattice data, though systematic errors, particularly for spin-2 states, affect the fit. The pseudo-scalar creates a massive resonance in the $0^{--}$ sector, and provides corrections to $0^{++}$ and spin 2 states. The mass has weak $N_c$ dependence, while the axion coupling shows minimal $N_c$ variation; see Ref~\cite{Athenodorou:2024loq}. The lattice data and theoretical predictions, shown in the left panel of Figure~\ref{fig:p0n1}, indicate higher energies for very long strings, possibly due to overestimation of heavy states' energy. Deviations in short string predictions may arise from lattice effects rather than effective string theory. 

The pseudoscalar is a metastable particle on the string world-sheet, which is described in the effective string theory by the action~\eqref{axion_interaction}. So we can now add massive excitations to the string. The spectrum of axionic states can be approximated by the $T\bar{T}$ dressing of free axions and undressed phonons, which works quantitatively well for most cases. 

Higher excited states, starting with the first excited state in $0^{--}$, are analyzed with more precise data, as shown in the right panel of Figure~\ref{fig:p0n1}. We apply this methodology to higher excited states, starting with the first excited state in the $0^{--}$ sector, using more precise data shown as dark red dots in the right panel of Figure~\ref{fig:p0n1}. This state can be computed using TBA, but higher-order corrections are needed due to the increased phonon momenta. By fitting the data using an expansion up to order $s^4 \ell_s^{-8}$ in the phase shift, a better fit for the spectrum is obtained (darker red solid line vs. dashed line). In the $0^{--}$ sector, another state below the GGRT level $N_L=N_R=2$ is identified, computed by solving the ABA and $T\bar{T}$ dressing equations. This state, shown as blue dots in the right panel of Figure~\ref{fig:p0n1}, is well-predicted by the $T\bar{T}$ dressing, with deviations at short string lengths likely due to higher-order corrections in the phase shift.

\begin{figure}[htb]
\scalebox{0.6}{{\includegraphics[width=0.72\textwidth]{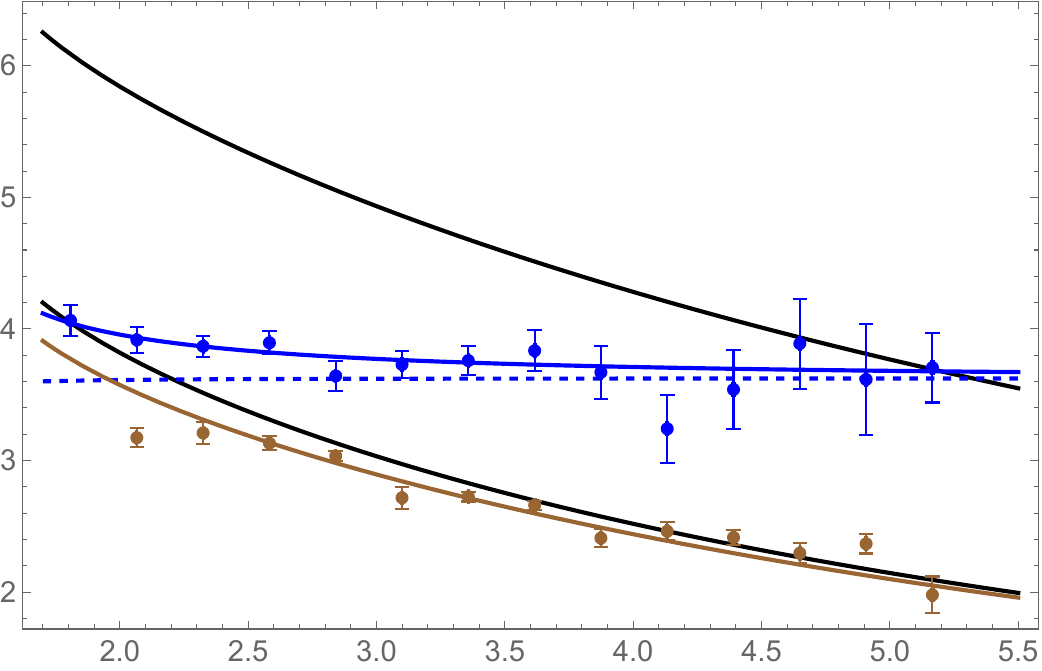}\put(-340,180){ \small $\Delta E\ell_s$}\put(0,0){\small $R/\ell_s$}}}
\centering
\caption{The $1^{\rm st}$ and $2^{\rm nd}$ excited states in the $0^{++}$ sector with $q=0$, labeled as brown and blue dots respectively. The solid and dashed blue lines represent the predictions of two massive excitations with and without $T\bar{T}$ deformation. The black lines are GGRT spectra at levels $N_L=N_R=1$ and $N_L=N_R=2$. The brown line is the prediction for the $N_L=N_R=1$ GGRT state with two phonon excitations, after introducing the PS term and axionic contribution.}
\label{fig:0pp_axion_axion}
\end{figure}

The second excited states in the $0^{++}$ sector are intriguing, especially the state with two axions, $A_0 A_0 |0\rangle$. As shown in Figure~\ref{fig:0pp_axion_axion}, the second excited state (blue dots) aligns well with two massive excitations after $T\bar{T}$ deformation (solid blue line), with a dashed blue line representing free axions for comparison. This suggests that the interaction between two pseudoscalars is significant and well captured by the $T\bar{T}$ deformation, particularly in the short string regime. The axion-axion interaction appears repulsive. 

\vspace{-0.75cm}
\section{The $q=1$ flux tube spectrum and the axion}
\label{sec:sectionq1}
\vspace{-0.25cm}

\begin{figure}[htb]
\scalebox{0.6}{{\includegraphics[width=0.65\textwidth]{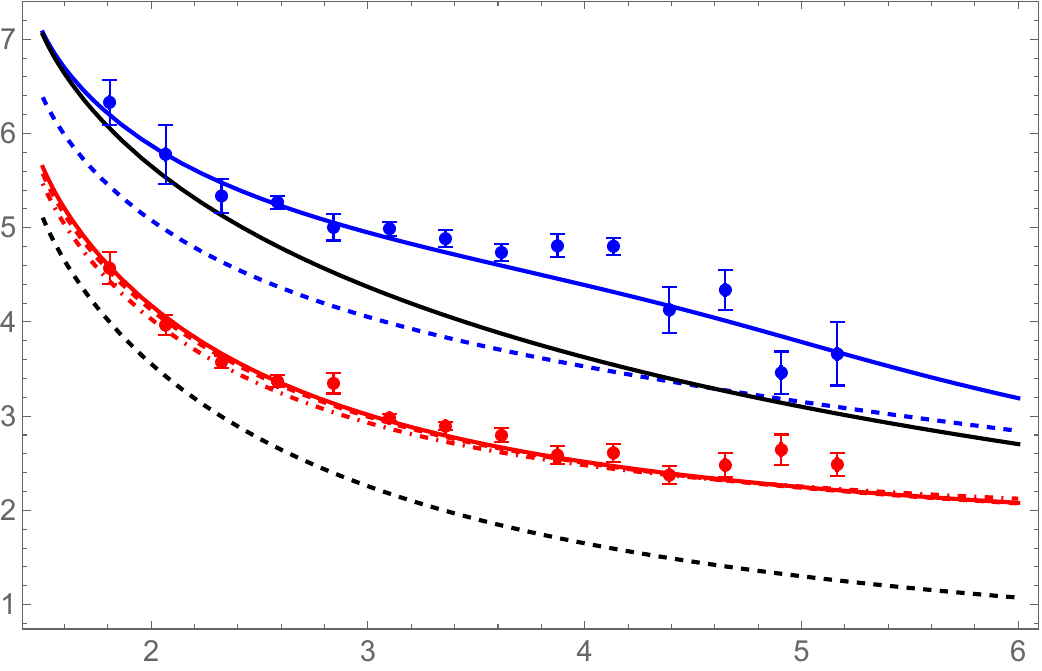}\put(-330,160){ \small $\Delta E\ell_s$}\put(0,0){\small $R/\ell_s$}} \hspace{1cm}
{\includegraphics[width=0.65\textwidth]{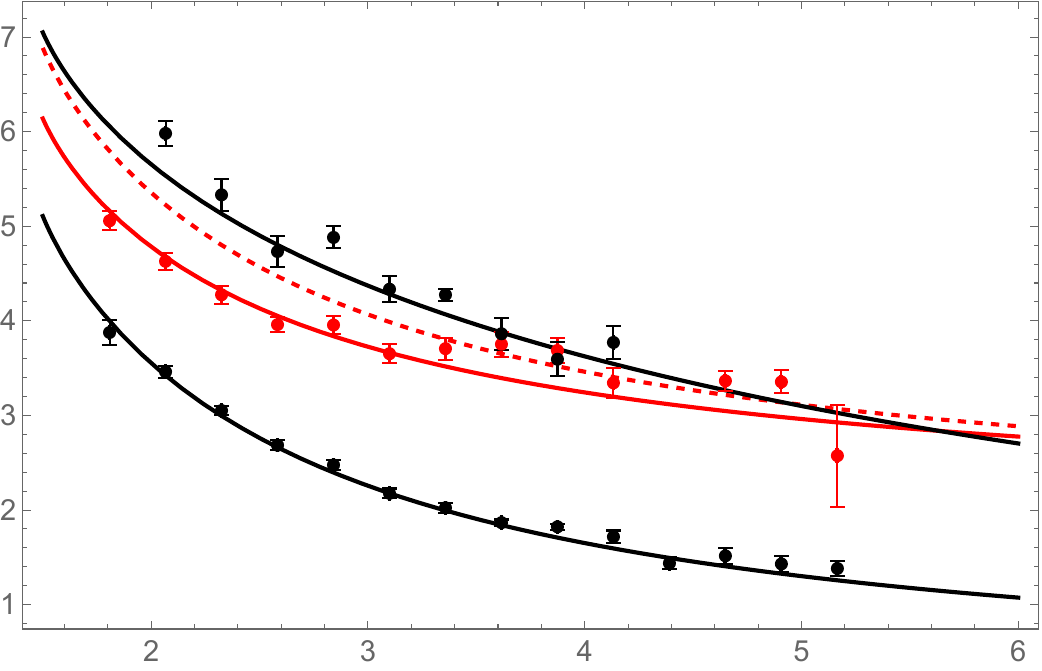}\put(-330,160){ \small $\Delta E\ell_s$}\put(0,0){\small $R/\ell_s$}}}
\centering
\caption{\underline{Left Panel}: Ground (red) and first excited states (blue) in the \(q = 1\); \(0^-\) sector. \underline{Right Panel}: The three lowest states in the \(q = 1\); \(1^{\pm}\) sector. Black dots represent the ground and second excited states, agreeing with GGRT spectra. Red dots show the first excited states, with theoretical predictions displayed by solid and dashed lines.}
\label{fig:q1}
\end{figure}

We first present results on the \(q = 1\); \(0^-\) sector, shown in left panel of Figure~\ref{fig:q1}. The ground state (red dots) can be interpreted either as a boosted \(2 \to 2\) phonon scattering in the pseudoscalar channel or as a boosted massive string excitation. We use ABA and \(T\bar{T}\) deformations for these interpretations, represented by solid and dot-dashed red lines. Both models predict the ground state well, with the two-phonon interpretation slightly more accurate. The first excited state, seen as a two-phonon excitation with quantization condition \(N_L = 2, N_R = 1\), shows poor agreement (dashed blue line), likely due to significant higher-order corrections at large phonon momenta. After including these corrections (solid blue line), predictions for the ground state remain unchanged, but the first excited state prediction improves significantly. We also calculate two-phonon scattering phase shifts from TBA equations which matches well with previously fitted phase shift, confirming that higher-order corrections are reliable. In the \(q = 1; 0^+\) sector, no accessible states with massive excitations are observed.

Next, in the \(q = 1\); \(1^{\pm}\) sector (Right panel in Figure~\ref{fig:q1}), data shows three states. The ground and second excited states (black dots) align well with the GGRT formula (black lines). Small deviations could be due to high momentum corrections. The first excited state (red dots) fits well with ABA and \(T\bar{T}\) deformation (solid red line). The \(T\bar{T}\) deformation models an attractive axion-phonon interaction but underestimates data for longer strings due to systematic errors. For short strings, \(T\bar{T}\) deformation gives accurate predictions.

For spin-2 states in the \(q = 1\) sector, data only covers the ground state, corresponding to the \(N_L = 2, N_R = 1\) GGRT state. 

\vspace{-0.8cm}
\section{Conclusion}
\label{sec:conclusions}

Our high-precision computations of the closed confining string spectrum align well with the theoretical predictions, which incorporate the low-energy effective string action and a low-mass worldsheet axion. We applied the $T\bar{T}$ deformation to describe the interactions, confirming no additional low-lying resonances on the string worldsheet up to the explored energy scale. However, some issues remain unresolved, such as the need for the development of a systematic approach to compute spectra with three or more phonon excitations. This would help disentangle multi-phonon states and extract non-universal Wilson coefficients relevant to 4D QCD flux tubes. The Axionic String Ansatz (ASA) was further supported as we found no evidence of extra resonances. 

\vspace{-0.4cm}
\section*{Acknowledgements}
\vspace{-0.4cm}
AA has been financially supported by the EuroCC2 project funded by the Deputy Ministry of Research, Innovation and
Digital Policy and the Cyprus Research and Innovation Foundation and the European High-
Performance Computing Joint Undertaking (JU) under grant agreement No 101101903 an well as by CaSToRC. SD and MT
acknowledge support by the Simons Collaboration on Confinement and QCD Strings. SD is also supported in part by the NSF grant PHY-2210349, and by  by the IBM Einstein Fellow Fund at the IAS.

\bibliographystyle{JHEP}
\bibliography{biblio_NEW}
\end{document}